\documentclass[11pt]{article}

\usepackage{amsfonts}
\usepackage{amsmath}
\usepackage{amssymb}
\usepackage{amsthm}

\usepackage{graphicx}

\theoremstyle{plain}
\newtheorem{definition}{Definition}
\newtheorem{theorem}{Theorem}

\theoremstyle{definition}

\DeclareMathOperator{\Int}{int}

\newcommand{\V}[1]{{\mathbf{#1}}}

\begin{document}

\title{On the reaction-diffusion replicator systems:\\ Spatial patterns and asymptotic behavior}

\author{Artem S. Novozhilov$^{1,}$\footnote{Corresponding author: anovozhilov@gmail.com}\,\,\,, Vladimir P. Posvyanskii$^1$,\\ and Alexander S. Bratus$^{1,2}$\\[3mm]
\textit{\normalsize $^{1}$Applied Mathematics--1, Moscow State University of Railway Engineering,}\\[-1mm]\textit{\normalsize Obraztsova 9, bldg. 9, Moscow 127994, Russia}\\[2mm]
\textit{\normalsize $^{2}$Faculty of Computational Mathematics and Cybernetics,}\\[-1mm]\textit{\normalsize Moscow State University, Moscow 119992, Russia}}

\date{}

\maketitle

\begin{abstract}
The replicator equation is ubiquitous for many areas of mathematical biology. One of major shortcomings of this equation is that it does not allow for an explicit spatial structure. Here we review analytical approaches to include spatial variables to the system. We also provide a concise exposition of the results concerning the appearance of spatial patterns in replicator reaction-diffusion systems.

\end{abstract}

\section{Replicator equation}\label{sec1}
The now classical replicator equation arises naturally in several important evolutionary contexts \cite{hofbauer1998ega,hofbauer2003egd,schuster1983rd}. The most straightforward way to write down the replicator equation is to start with a \textit{symmetric two player game} with a payoff matrix $\V{A}=(a_{ij})_{k\times k}$, where $a_{ij}$ denotes the payoff of player one if player one uses pure strategy $i$ against pure strategy $j$ used by player two. Let $\V{p}=(p_1,\ldots,p_k)$ denote a mixed strategy, which belongs to the simplex $S_k=\{\V{p}\in\mathbb R^k\colon \sum_i p_i=1,\,p_i\geq 0,\,i=1,\ldots,k\}$. The average payoff of pure strategy $i$ against $\V{p}$ is $(\V{Ap})_i=\sum_j a_{ij}p_j$, and the average payoff of mixed strategy $\V{q}$ against $\V{p}$ is $\langle\V{q},\V{Ap}\rangle=\sum_{i}q_i(\V{Ap})_i=\sum_{ij}a_{ij}p_iq_j$.

For the following we will need the notion of the Nash equilibrium. The \textit{Nash equilibrium} of the game with the payoff matrix $\V{A}$ is defined as such strategy $\V{\hat{p}}\in S_k$ that fares best against itself, i.e.
\begin{equation}\label{eq1:1}
    \langle \V{p},\V{A\hat{p}}\rangle\leq \langle\V{\hat{p},\V{A\hat{p}}}\rangle,
\end{equation}
for any $\V{p}\in S_k$.

Now let us identify the payoff of a game with an individual's fitness and instead of two players consider a populations of players, each of which can use one out of $k$ different strategies (these strategies can be mixed in general, but at this point it is more transparent to assume that the strategies are pure). The reproductive success of an individual using strategy $i$ depends on its average payoff $(\V{Ap})_i$ assuming random encounters with other individuals. Here $\V{p}\in S_k$ is the current structure of the population, $p_i=n_i/N$, where $n_i$ is the absolute number of players using strategy $i$ and $N=\sum_j n_j$ is the total population size. We obtain, in a usual manner, that
\begin{equation}\label{eq1:2}
\frac{\dot{n}_i}{n_i}=\V{(Ap)}_i,\quad i=1,\ldots,k.
\end{equation}
System \eqref{eq1:2} can be called a selection system, according to the general terminology \cite{karev2009,karev2010}. If instead of absolute sizes we consider the corresponding frequencies, system \eqref{eq1:2} reduces to
\begin{equation}\label{eq1:3}
    \dot p_i=p_i((\V{Ap})_i-\langle\V{p},\V{Ap}\rangle),\quad i=1,\ldots,k.
\end{equation}
Equation \eqref{eq1:3} in this general form was written down first in \cite{taylor1978ess} and was dubbed ``replicator equation'' in \cite{schuster1983rd}; particular cases of \eqref{eq1:3} were considered previously in the population genetics \cite{hofbauer1998ega,svirezhev:fme} and in the mathematical theories of prebiotic molecular evolution \cite{bnp2010,eigen1971sma,eigen1989mcc,eigenshuster}. Note that if a solution to system \eqref{eq1:2} is known then the corresponding solution to \eqref{eq1:3} can be calculated. In general different selection systems of the form \eqref{eq1:2} correspond to the replicator equation \eqref{eq1:3}; e.g., instead of $(\V{Ap})_i$ one can consider the fitness of the form $(\V{Ap})_i+F(N)$, where $F$ is an arbitrary function (see also \cite{karev2010}).

There exists a parallel between the concepts of game theory with the payoff matrix $\V{A}$ and the behavior of the solutions to the replicator equation \eqref{eq1:3}. For example, if $\V{\hat{p}}\in S_k$ is a Nash equilibrium then $\V{\hat{p}}$ is a stationary point of \eqref{eq1:3}. If $\V{\hat{p}}$ is a stationary point of \eqref{eq1:3} and Lyapunov stable then $\V{\hat{p}}$ is a Nash equilibrium.

Another important notion of the game theory is that of \textit{evolutionary stable strategy} (ESS). Strategy $\V{\hat{p}}$ is called ESS if \eqref{eq1:1} holds for any $\V{p}\in S_k$ and, additionally, for all $\V{p}\neq \V{\hat{p}}$ such that $\langle \V{p},\V{A\hat{p}}\rangle=\langle\V{\hat{p},\V{A\hat{p}}}\rangle$ it is true that $\langle \V{p},\V{A{p}}\rangle< \langle\V{\hat{p},\V{A{p}}}\rangle$. It can be proved that the last two conditions are equivalent to
\begin{equation}\label{eq1:4}
    \langle \V{\hat{p}},\V{A{p}}\rangle> \langle\V{{p},\V{A{p}}}\rangle
\end{equation}
for all $\V{p}\in S_k$ from some neighborhood of $\V{\hat{p}}$. With some abuse of notation a stationary point $\V{\hat{p}}$ of \eqref{eq1:3} is called an \textit{evolutionary stable state} (and abbreviated the same ESS) if \eqref{eq1:4} holds; being ESS implies that $\V{\hat{p}}$ is asymptotically stable (the opposite is not true).

Equation \eqref{eq1:3} is unrealistic in that it ignores the population structure, all the individuals using the same strategies are supposed to be identical, the population itself being well mixed. The population structure can have different nature, e.g., the contact structure, age structure, physiological structure, etc \cite{Novozhilov2008b,Nowak2010}. One extremely important structural variable is the space \cite{ssc2000}. Among different modeling approaches to include the spatial structure in the model, reaction-diffusion equations take an important part \cite{cantrell2003spatial}.
\section{Reaction-diffusion replicator equation}\label{sec2}
The usual tactics to consider mathematical models with explicit spatial structure is to replace a system of ordinary differential equations $$\dot{\V{n}}=\V{f(n)}$$ with the system of partial differential equations of the form $$\partial_t{\V{n}}=\V{f(n)}+\V{D}\Delta \V{n},$$ where $\V{D}$ is a diagonal matrix of diffusion coefficients and $\Delta$ is the Laplace operator. This approach however fails in general with the replicator equation~\eqref{eq1:3} because there is an additional condition that $\V{p}\in S_k$. The only case when this condition can be met is when all the diffusion coefficients are identical (this case is a generalization of the classical work \cite{fisher1937waa}). In this case we obtain the following system of reaction diffusion equations
\begin{equation}\label{eq2:1}
    \partial_t p_i=p_i((\V{Ap})_i-\langle\V{p},\V{Ap}\rangle)+d\Delta p_i,\quad i=1,\ldots,k,
\end{equation}
that was studied in \cite{hadeler1981dfs} together with the boundary conditions $\partial_\V{l}p_i=0$, where $\V{l}$ is the normal to the boundary of the domain in which the solution to \eqref{eq2:1} is sought, we denot this domain $\Omega$.

The following main result concerning \eqref{eq2:1} was proved in \cite{hadeler1981dfs}:
\begin{theorem}\label{th2:1}
Let system \eqref{eq1:3} have an inner asymptotically stable stationary point $\V{\hat{p}}\in\Int S_k=\{\V{p}\in S_k\colon \V{p}>0\}$. Then $\V{\hat{p}}$ is a globally asymptotically stable stationary point of \eqref{eq2:1} for any diffusion coefficient $d$.
\end{theorem}
There are several possible approaches to study the replicator equation with explicit space with different diffusion coefficients. The first approach by Vickers and co-authors uses the principle of local population regulation such that the fitness values are adjusted in such a way so that the total local population size stays constant \cite{ferriere2000ads,hutson1992travelling,hutson1995sst,vickers1989spa,Vickers1991,vickers1993spatial}. To be more precise, as a counterpart of \eqref{eq1:3} Vickers et al. consider the system
\begin{equation}\label{eq2:2}
    \partial_t n_i=n_i((\V{Ap})_i-\langle\V{p},\V{Ap}\rangle)+d_i\Delta p_i,\quad i=1,\ldots,k,
\end{equation}
which is written for the absolute sizes $n_i(\V{x},t)$, where $\V{x}\in\Omega\subset\mathbb R^m,\,m=1,2,3$. System \eqref{eq2:2} should be supplemented with the initial and boundary conditions.

Another approach by Cressman and co-authors \cite{cressman1987density,cressman1997sad,prior1993evolutionary} is to consider system \eqref{eq1:2} first and then add the space. This approach also uses absolute sizes for the subpopulations using different strategies:
\begin{equation}\label{eq2:3}
    \partial_t n_i=n_i((\V{Ap})_i+F(N))+d_i\Delta p_i,\quad i=1,\ldots,k,
\end{equation}
where $F$ is typically a smooth decreasing function. A similar approach to consider ``open'' replicator systems is discussed in \cite{bratus2009stability}.

Both systems \eqref{eq2:2} and \eqref{eq2:3} are equivalent to the replicator equation \eqref{eq1:3} when there is no spatial structure, however they have remarkably different behaviors for $d_i>0$ (see below).

Finally, another approach was used in \cite{bnp2010,bratus2006ssc,bratus2009existence,bratus2011,weinberger1991ssa} to study a spatially explicit counterpart of the replicator equation \eqref{eq1:3}. In short, we start with the system for the absolute sizes
\begin{equation*}
    \dot n_i=n_i(\V{Ap})_i+d_i\Delta p_i,\quad i=1,\ldots,k,
\end{equation*}
and use the transformation $p_i=n_i/\sum_j \int_\Omega n_j(\V{x},t)\,d\V{x}$ to obtain
\begin{equation}\label{eq2:4}
    \partial_tp_i=p_i \left[(\V{Ap})_i-{f}^{sp}(t)\right]+d_i\Delta p_i,\quad i=1,\ldots,k.
\end{equation}
Here $p_i=p_i(\textbf{x},t),\,\textbf{x}\in\Omega,\,\partial_t=\frac{\partial}{\partial t}\,,\,\Delta$ is the Laplace operator, and $d_i>0,\,i=1,\ldots,k$ are the diffusion coefficients. The initial conditions are $p_i(\V{x},0)=\varphi_i(\V{x}),\,i=1,\ldots,k,\,$ and
\begin{equation}\label{eq2:5}
    {f}^{sp}(t)=\sum_{i}\int_{\Omega}p_i(\V{Ap})_i\,d\textbf{x}=\int_{\Omega}\langle \V{p},\,\V{Ap}\rangle\,d\textbf{x}.
\end{equation}
For the system \eqref{eq2:4} we have the condition of the global regulation
\begin{equation}\label{eq2:6}
\sum_{i} \int_{\Omega}p_i(\textbf{x},t)\,d\textbf{x}=1.
\end{equation}
Solutions to \eqref{eq2:4} now belong to the analog of the simplex $S_k(\Omega_t),\,\Omega_t=\Omega\times[0,\infty)$ and for any time moment $t$ are assumed to belong to the Sobolev space $W_2(\Omega)$ and satisfy condition \eqref{eq2:6}.

Formally, system \eqref{eq2:4} can be obtained from \eqref{eq2:3} if we assume that $N=N(t)=\sum_i\int_{\Omega}n_i(\V{x},t)\,d\V{x}$, therefore the results for systems \eqref{eq2:3} and \eqref{eq2:4} should be similar.

We are mostly interested in the steady state solutions to the replicator reaction-diffusions systems. Of especial interest is the stability of spatially homogeneous stationary points, which are also stationary solutions to the replicator equation \eqref{eq1:3}. For systems \eqref{eq2:2} and \eqref{eq2:3} it is possible to study also traveling wave solutions, but these particular kinds of solutions are not considered here (see for an extensive discussion, e.g., \cite{cressman1997sad,Ferriere1995,ferriere2000ads,hofbauer1997travelling,hutson1992travelling,Vickers1991}).

\section{Turing instabilities and spatial patterns in replicator reaction-diffusion systems}\label{sec3}
Let us assume that $\V{\hat{p}}\in\Int S_k$ is an ESS for the game with payoff matrix $\V{A}$. Then it is an asymptotically stable state of the replicator equation \eqref{eq1:3}. This stationary point is also a spatially homogeneous stationary solution for the systems \eqref{eq2:3}, \eqref{eq2:4} (with some choice of $N=\hat{N}$) and \eqref{eq2:5}. Will this solution be asymptotically stable for the spatially explicit systems?

For the system \eqref{eq2:2} Vickers \cite{vickers1989spa} proved the following theorem:
\begin{theorem}\label{th3:1}
Let $\V{\hat{p}}$ be an ESS of the game with the payoff matrix $\V{A}$. Then $\V{\hat{n}}=\hat{N}\V{\hat{p}}$ is an asymptotically stable spatially homogeneous stationary solution to \eqref{eq2:2} for any diffusion coefficients.
\end{theorem}
If $\V{\hat{p}}\in\Int S_k$, Theorem \ref{th3:1} is a generalization of Theorem \ref{th2:1}. The major difference of system \eqref{eq2:3} is that an ESS can become unstable for the system with the space. In general, it holds that
\begin{theorem}\label{th3:2}
Let $\V{\hat{p}}$ be an ESS of the game with the payoff matrix $\V{A}$. Then $\V{\hat{n}}=\hat{N}\V{\hat{p}}$ is a spatially homogeneous solution to \eqref{eq2:3}, which can become unstable in spatially explicit system for a particular choice of $\V{A}$ and diffusion coefficients.
\end{theorem}
Several theorems on the behavior of the ESSs, which specify exact conditions on the appearance of the spatial patterns for particular cases $k=2,3$ can be found in \cite{cressman1997sad}.

Using system \eqref{eq2:4} it is possible to give more exact general statements concerning appearance of the spatially inhomogeneous solutions \cite{bratus2006ssc,bratus2009existence,bratus2011}. In particular, the following necessary condition was proved in \cite{bratus2011}.

\begin{theorem}\label{th3:3} Let $\V{\hat{p}}\in\Int S_n$ be an asymptotically stable rest point of \eqref{eq1:3}. Then for this point to be an asymptotically stable homogeneous stationary solution to \eqref{eq2:4} it is necessary that
\begin{equation}\label{eq3:1}
    \sum_{i} d_i> \frac{\beta}{\lambda_1}\,,\quad \beta=\langle\V{A\hat{p}},\,\V{\hat{p}}\rangle,
\end{equation}
where $\lambda_1$ is the first non-zero eigenvalue of the following eigenvalue problem
\begin{equation*}
    \Delta \psi(\V{x})+\lambda \psi(\V{x})=0,\quad \V{x}\in\Omega,\quad \partial_\V{l}\psi|_{\textbf{x}\in\Gamma}=0.
\end{equation*}
\end{theorem}
Theorem \ref{th3:3} points out that to expect an appearance of spatially inhomogeneous solutions the diffusion coefficients have to be sufficiently small. For some particular choices of $\V{A}$ it is possible to give an exact conditions for the appearance of spatial patterns \cite{bratus2006ssc,bratus2009existence}. In particular, consider matrix $\V{A}$ of the form
$$
\V{A}=\begin{pmatrix}
        a_1 & 0 & \ldots & 0 \\
        0 & a_2 & \ldots & 0 \\
        \ldots &  &  &  \\
        0 & \ldots & 0 & a_k \\
      \end{pmatrix},
$$
then (see \cite{bratus2009existence})
\begin{theorem}\label{th3:4} There exists a spatially non-uniform stationary solution to \eqref{eq2:4} with the given $\V{A}$ in case of $\Omega=[0,1]$ if
$$
\sum_i \frac{d_i}{a_i}<\frac{1}{\pi^2}\,.
$$
\end{theorem}
Moreover, using the special hamiltonian form of the equations for the stationary solutions to \eqref{eq2:4} it is possible to obtain a parametric form for these spatially non-uniform solutions \cite{bratus2006ssc,bratus2009existence}.
\section{Distributed Nash equilibrium and evolutionary stable state (DESS)}
Since the appearance of spatially heterogeneous solutions in system \eqref{eq2:4} is a generic phenomenon, it is necessary to study the asymptotic behavior of the distributed system, especially given the fact that for some values of the diffusion coefficients there are no stable spatially homogeneous solutions in the system. We approached this problem by considering a generalization of the notions of the Nash equilibria and ESS to the spatially explicit case \cite{bratus2011}.

\begin{definition}\label{def4:1}
We shall say that the vector function $\V{\hat{w}}(\V{x})\in S_k(\Omega)$ is a distributed Nash equilibrium if
\begin{equation}\label{eq4:1}
    \int_{\Omega} \langle \V{u}(\V{x},t),\,\V{A\hat{w}}(\V{x})\rangle\, d\V{x}\leq \int_{\Omega}\langle\V{\hat{w}}(\V{x}),\,\V{A\hat{w}}(\V{x})\rangle \,d\V{x}
\end{equation}
for any vector-function $\V{u}(\V{x},t)\in S_k(\Omega_t),\,\V{u}(\V{x},t)\neq \V{\hat{w}}(\V{x})$.
\end{definition}

\begin{definition}
We shall say that $\V{\hat{w}}(\V{x})\in S_k(\Omega)$ is a distributed evolutionary stable state \emph{(}DESS\emph{)} if
\begin{equation}\label{eq4:2}
\int_{\Omega}\langle \V{\hat{w}}(\V{x}),\,\V{Au}(\V{x},t)\rangle d\V{x}>\int_{\Omega}\langle \V{u}(\V{x},t),\,\V{Au}(\V{x},t)\rangle\,d\V{x}
\end{equation}
for any $\V{u}(\V{x},t)\in S_k(\Omega_t)$ from a neighborhood of $\V{\hat{w}}(\V{x})\in S_k(\Omega)$, $\V{u}(\V{x},t)\neq \V{\hat{w}}(\V{x})$.
\end{definition}

We also need the following stability notion
\begin{definition}\label{def3:2}
Stationary solution $\V{\hat{w}}(\V{x})\in S_k(\Omega)$ to \eqref{eq2:4} is stable in the sense of the mean integral value if for any $\varepsilon>0$ there exists $\delta>0$ such that for the initial data $\varphi_i(\bf{x})$ of system \eqref{eq2:4}, which satisfy
$$
|\bar{\varphi}_i-\hat{\bar{{w}}}_i|<\delta,\quad i=1,\ldots,n,
$$
where
$$
\bar{\varphi}_i=\int_\Omega\varphi_i(\V{x})\,d\V{x},\quad \hat{\bar{w}}_i=\int_\Omega \hat{w}_i(\V{x})\,d\V{x},
$$
it follows that
$$
|\bar{u}_i(t)-\hat{\bar{w}}_i|<\varepsilon
$$
for any $i=1,\ldots,n$ and $t>0$.

Here $\bar{u}_i(t)=\int_\Omega u_i(\V{x},t)\,d\V{x}$ and $u_i(\V{x},t),\,i=1,\ldots,n$ are the solutions to \eqref{eq2:4}, $\V{u}(\V{x},t)\in S_k(\Omega_t)$.
\end{definition}

Using these definitions we can prove the following theorems \cite{bratus2011}
\begin{theorem}\label{th4:1}
If $\V{\hat{w}(x)}\in\Int S_k(\Omega)$ is a Lyapunov stable stationary solution to \eqref{eq2:4} then $\V{\hat{w}}(\V{x})$ is a distributed Nash equilibrium.
\end{theorem}

\begin{theorem}\label{th4:2}
Let $\V{\hat{w}}\in\Int S_k$ be a spatially homogeneous solution to \eqref{eq2:4} \emph{(}i.e., $\V{\hat{w}}\in\Int S_k$ is a rest point of \eqref{eq1:3}\emph{)}. If $\V{\hat{w}}$ is DESS then $\V{\hat{w}}$ is an asymptotically stable solution to \eqref{eq2:4} in the sense of the mean integral value.
\end{theorem}

In words, Theorem \ref{th4:2} implies that if $\V{\hat{p}}\in S_k$ is an ESS, then the corresponding spatially homogeneous solution will stay stable, albeit in the weaker mean integral sense if $\V{\hat{p}}$ is DESS. The sufficient condition of being DESS is given by
\begin{theorem}\label{th4:3}
The stationary point $\V{\hat{p}}\in\Int S_k$ is a DESS if
\begin{equation}\label{eq5:6}
\langle \V{Ac}^0(t),\V{c}^0(t)\rangle\leq -\gamma^2|\V{c}^0(t)|^2,\quad \gamma>0
\end{equation}
for any $\V{c}^0(t)$ satisfying $\sum_i c_i^0(t)=0$.
\end{theorem}
Proof of Theorem \ref{th4:3} can be found in \cite{bratus2011} together with some examples.
\section{Discussion}
Here we give a short review of the analytical work on the replicator reaction-diffusion equations. There are different approaches to include the explicit space in the replicator equation \eqref{eq1:3}. One of them, by Vickers et al., supposes that there is local regulation of the population. While this approach yields simple results (see Theorem \ref{th3:1}) concerning the appearance of spatial patterns, it is difficult to assume that the principle of local regulation is abundant in real systems. The approach used by Cressman and co-authors (equation \eqref{eq2:3}) \cite{cressman1987density,cressman1997sad} and our approach (equation \eqref{eq2:4}) both imply the principle of global regulation whereas the total population size is being regulated by internal or external means (see also \cite{weinberger1991ssa} for discussion). The major difference between the local and global regulation approaches appears when a fate of an ESS of the local replicator system is studied. While in system \eqref{eq2:2} this stationary point remains stable, in systems \eqref{eq2:3} and \eqref{eq2:4} this point can loose its stability through Turing's mechanism. The analysis of system \eqref{eq2:4} shows that this is probable when the diffusion coefficients are sufficiently small.

The approach, adopted in our works \cite{bratus2006ssc,bratus2011}, allows for further generalizations of the notions of the Nash equilibria and ESS. In general, we can conjecture that the behavior of the distributed replicator equation of the form \eqref{eq2:4} mimics that of the local system \eqref{eq1:3} if the stability notion is replaced with a weaker notion of the stability in the mean integral sense.

The reviewed approaches are not the only possible ways to study replicator equations with explicit space. A lot of work was devoted to the study of particular class of replicator equations arising in the area of mathematical models of prebiotic evolution \cite{bratus2009existence,eigenshuster}. In particular, appearance of spiral ways was studied by means of numerical simulations of some partial differential equations \cite{cronhjort1995hvp,cronhjort2000ibr,cronhjort1994hvp,cronhjort1997ccm,cronhjort1996hhn} and cellular automata \cite{boerlijst1990ssa,boerlijst2000sas,boerlijst1991sws,boerlijst1995sge,boerlijst1993ecs}. These approaches are complementary to those considered in this short note. It is also possible to consider evolutionary systems with discrete state space (e.g., \cite{blume1993statistical,ellison1993learning,kosfeld2002stochastic}) but in general the better elaborated theory of reaction-diffusion equations allows deaper analytical insights into the behavior of replicator equations with explicit spatial structure.

\paragraph{Acknowledgments:} The research is supported in part by the Russian Foundation for Basic Research grant \# 10-01-00374. ASN is supported by the grant to young researches from Moscow State University of Railway Engineering.


\end{document}